\documentclass[aps,prl,superscriptaddress,reprint,longbibliography,floatfix]{revtex4-2}
\usepackage{graphicx}
\usepackage{amsmath}
\usepackage{amssymb}
\usepackage{bm}
\usepackage{color}
\usepackage[utf8]{inputenc}
\usepackage[T1]{fontenc}
\usepackage{newunicodechar}
\usepackage{xr-hyper}
\usepackage[colorlinks=true, allcolors=blue]{hyperref}
\usepackage{xcolor}
\usepackage{kotex}
\usepackage{braket}
\usepackage{booktabs}

\newcommand{\dd}{\mathrm d}
\newcommand{\ii}{\mathrm i}
\newcommand{\ee}{\mathrm e}
\newcommand{\avg}[1]{\left\langle #1\right\rangle}
\newcommand{\abs}[1]{\left|#1\right|}
\newcommand{\vq}{\bm q}
\newcommand{\br}{\bm r}
\newcommand{\vV}{\bm V}

\newcommand{\vu}{\bm u}
\newcommand{\vn}{\bm n}
\newcommand{\vF}{\bm F}

\newcommand{\order}{\mathcal O}

\begin{document}
\title{Trajectory Statistics Govern Mechanical Power Transfer in Active Baths}
\author{Chul-Ung Woo}
\email{chul-ung.woo@uni-saarland.de}
\affiliation{Department of Theoretical Physics and Center for Biophysics, Saarland University, Saarbr\"ucken, Germany}
\author{Jiwon Choi}
\email{chlwldnjs175@gmail.com}
\affiliation{Department of Theoretical Physics and Center for Biophysics, Saarland University, Saarbr\"ucken, Germany}
\author{Heiko Rieger}
\email{heiko.rieger@uni-saarland.de}
\affiliation{Department of Theoretical Physics and Center for Biophysics, Saarland University, Saarbr\"ucken, Germany}
\date{\today}

\begin{abstract}
In active media, a moving passive object can experience negative drag and receive positive mechanical power from the bath, in contrast to the dissipative response in equilibrium.
For a probe moving at velocity $\vV$ and interacting weakly with an ideal active bath, we show that free-particle trajectory statistics determine the density response and the resulting bath force.
The probe--particle interaction sets the wave-vector weighting, while the imposed motion selects frequencies $\omega_{\vq}=\vq\cdot\vV$, separating bath dynamics from probe properties.
The sign of the response spectrum yields probe-independent constraints for mechanical power transfer.
Applying this framework to representative active-particle models reveals distinct power-transfer behavior despite identical displacement variances.
For prescribed probe motion, the relation predicts drag reversal, motion-induced density distortions, and velocity-dependent friction, in quantitative agreement with simulations.
Beyond the mean force, the free-particle trajectory statistics also determine the force autocorrelation.
These results provide a route to predict the mechanical response of probes from single-particle trajectory statistics measured in the absence of the probe.
\end{abstract}

\maketitle

%\textit{Introduction---}
%Passive objects in active media experience unusual transport and forces arising from the motion of surrounding active particles [1–3]. Across microbial suspensions, motile-cell carpets, bacterial turbulence, and active granular media, tracers exhibit enhanced diffusion, non-Gaussian displacements, transient superdiffusion, and persistent memory [4–11]. These transport effects have been linked to swimmer–tracer encounters, fluctuating active forces, and crowding [12–15]. Beyond tracer transport, active particles exert pressure on boundaries [16, 17] and mediate interactions between passive bodies [18–21]. Active baths can also drive directed motion of asymmetric objects [22–27]

Passive objects immersed in active media exhibit unusual transport phenomena and experience forces generated by the motion of surrounding active particles~\cite{chen2007fluctuations,bechinger2016active,granek2024colloquium}.
These effects manifest as enhanced diffusion, non-Gaussian displacement statistics, transient superdiffusion, and persistent memory across a wide range of active systems, from microbial suspensions and motile-cell carpets to bacterial turbulence and active granular media~\cite{wu2000particle,leptos2009dynamics,mino2011enhanced,jepson2013enhanced,jeanneret2016entrainment,grossmann2024non-gaussian,xie2022activity-induced,caprini2024emergent}.
The underlying mechanisms have been linked to swimmer–tracer encounters, fluctuating active forces, and crowding~\cite{burkholder2017tracer,dhar2024active,reichert2021tracer,granek2022anomalous}.
Active particles also exert boundary pressures with properties distinct from those of passive baths~\cite{takatori2014swim,solon2015pressure}, mediate interactions between passive objects~\cite{ni2015tunable,baek2018generic,granek2020bodies,ning2023hydrodynamics} and are able to drive the directed motion of asymmetric objects~\cite{angelani2009selfstarting,sokolov2009swimming,dileonardo2010bacterial,mallory2014curvature-induced,pietzonka2019autonomous,pellicciotta2025wall}.

A different response arises for a passive probe translated at a prescribed velocity in an active bath: the bath can assist rather than resist its motion, producing negative drag and transferring positive mechanical power~\cite{foffano2012colloids,knippenberg2025negative}.
Unlike the directed motion of asymmetric objects, this response can occur for symmetric probes in inversion-symmetric baths~\cite{kim2024symmetry-breaking}.
Determining whether the bath assists or resists a given probe motion thus requires relating active-particle dynamics and the probe interaction to the net force.

To understand this force, microrheological studies have examined the response of active particles to a moving probe~\cite{reichhardt2015active,burkholder2020nonlinear,peng2022forced,john2025progress}.
Related theories describe density distortions, friction, fluctuations, and memory~\cite{maes2015friction,demery2014generalized,maes2020fluctuating,davis2024active,pei2025induced,burkholder2019fluctuation,solon2022einstein}, as well as effective interactions and responses to external potentials~\cite{steffenoni2016interacting,dalcengio2019linear,kafri2026active}.
For one-dimensional run-and-tumble particle (RTP) and two-dimensional active Brownian particle (ABP) baths, recent calculations have obtained velocity-dependent friction and noise~\cite{pei2026transfer}.
In many such approaches, the force is expressed using stationary distributions or dynamical correlations conditioned on the probe configuration or motion~\cite{maes2015friction,maes2020fluctuating,pei2025induced,pei2026transfer}.

In equilibrium linear response theory relates the drag coefficient
of a probe that moves with constant velocity through a liquid 
via a Green-Kubo formula to the equilibrium force correlations \cite{kubo1966fluctuation}. In active baths these 
relations need not hold~\cite{dalcengio2019linear}, since no equilibrium states can serve as reference unperturbed states.
In principle, the force correlations could be expressed with the help of the free propagator of the bath particles, which allows to relate the drag coefficient to the dynamics of the bath particles alone without referring to a stationary state or a specific probe. 
This approach has not been followed up to now, although it appears particularly promising for active baths, for which reason we will present it here.

We find that in the weak probe limit what is needed to determine its mechanical response is not only the displacement fluctuations of the active particles but the full distribution of active particle displacements at each time lag, including higher-order statistics beyond the variance. This quantity is commonly denoted as the self-intermediate scattering function (self-ISF), the spatial Fourier transform of the single-particle displacement distribution~\cite{hoefling2013anomalous,kurzthaler2016intermediate}.

\begin{figure*}[t]
\centering
\includegraphics[width=0.98\textwidth]{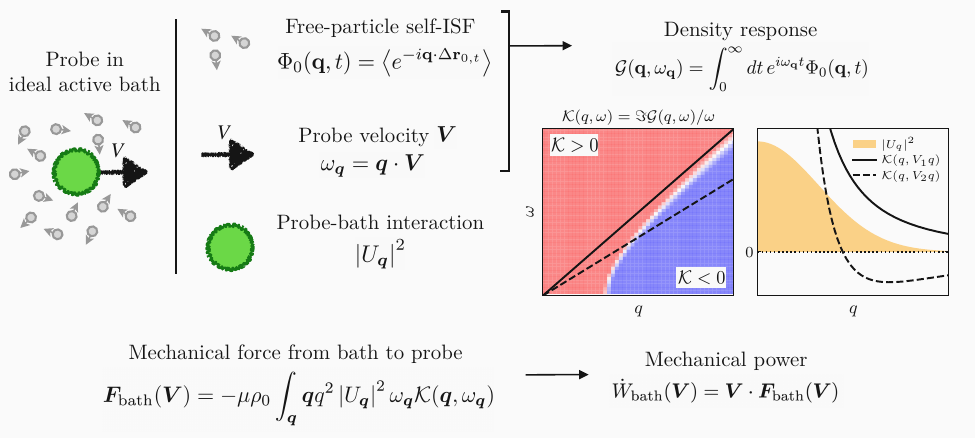}
\caption{\textbf{From free-particle motion to the mechanical response of a moving probe.}
The moving-probe problem separates into a bath response determined by the free-particle self-ISF, frequency sampling set by the probe velocity, and wave-vector weighting set by the probe-bath interaction.
The right panels illustrate the sampled responses for two probe velocities together with the interaction weight. 
Combining them in the force integral gives the bath force and the mechanical power transferred from the bath to the probe.}
\label{fig:overview}
\end{figure*}

We will show how this statistics relates to the force on a weak moving probe by deriving an explicit relation between the free-particle self-ISF and the density and force responses of a weak, uniformly translating probe in an ideal active bath [Fig.~\ref{fig:overview}].
This relation yields probe-independent power-transfer constraints and probe-dependent predictions for drag and density and specifically this general framework allows the formulation of various theorems on the possibility of, for instance, negative drag for specific active particles based on their displacement statistics alone.

\textit{Trajectory-response relation---}
A probe moves at constant velocity $\vV$ through a bath of noninteracting active particles of far-field density $\rho_0$.
The active particle dynamics within the probe frame is given by
\begin{equation}
\dot{\br}=\vu(s_t)-\vV-\mu\nabla U(\br)+\sqrt{2D_t}\,\bm\eta(t),
\label{eq:model}
\end{equation}
where $s_t$ is a stationary Markov propulsion process with velocity $\vu(s_t)$ and generator $\mathcal L_s$; 
$\mu$ is the mobility, $D_t$ the translational diffusivity, and $\bm\eta$ independent unit Gaussian white noise.
The probe couples to the active bath 
through a localized potential $U(\br)$, where $\br$ is the particle position relative to the probe.
Writing the stationary density as $\rho=\rho_0+\delta\rho$, the 
force exerted by the bath on the probe is then 
\begin{equation}
\vF_{\rm bath}(\vV)=\int\dd^dr\,\delta\rho(\br;\vV)\nabla U(\br),
\label{eq:force_density_perturbation}
\end{equation}
since $\int\dd^dr\,\nabla U=0$.
Linearizing the density dynamics in $U$ and propagating the perturbation with the free dynamics yields the density response in terms of free-particle displacement statistics (see SM~\cite{prlsm}).

Let $\Delta\br_{0,t}$ denote the free-particle displacement
in the laboratory frame, and define the self-ISF $\Phi_0(\vq,t)=\avg{\ee^{-\ii\vq\cdot\Delta\br_{0,t}}}$, the Fourier transform of the free-particle displacement distribution.
The corresponding frequency-dependent response function is
\begin{equation}
\mathcal G(\vq,\omega)=\int_0^\infty\dd t\,\ee^{\ii\omega t}\Phi_0(\vq,t).
\label{eq:trajectory_response_characteristic}
\end{equation}
We use the Fourier convention $f(\br)=\int_{\vq}\ee^{\ii\vq\cdot\br}f_{\vq}$, with $\int_{\vq}\equiv\int\dd^dq/(2\pi)^d$.
Uniform translation contributes the phase $\ee^{\ii\omega_{\vq}t}$, with $\omega_{\vq}=\vq\cdot\vV$, so each wave vector samples the response at a frequency set by the imposed velocity.
The resulting density response is
\begin{equation}
\delta\rho_{\vq}(\vV)=-\mu\rho_0q^2\mathcal G(\vq,\omega_{\vq})U_{\vq}.
\label{eq:density_susceptibility}
\end{equation}
Equation~\eqref{eq:density_susceptibility} has the standard linear-susceptibility form~\cite{kubo1966fluctuation,agarwal1972fluctuation,marconi2008fluctuation,baiesi2009fluctuations,dalcengio2019linear}.
In the present ideal bath, however, the susceptibility is determined entirely by the probe-free self-ISF, without first solving for the probe-conditioned stationary distribution.

The leading-order bath force is then given by:
\begin{equation}
F_{{\rm bath},i}(\vV)=-\mu\rho_0\int_{\vq}q_iq^2|U_{\vq}|^2
\Im\mathcal G(\vq,\omega_{\vq}).
\label{eq:finiteV_general}
\end{equation}
Thus bath trajectories determine $\mathcal G$, the interaction sets the nonnegative weight $|U_{\vq}|^2$, and the imposed velocity selects $\omega_{\vq}$~\footnote{Using $U_{-\vq}=U_{\vq}^{*}$ and $\mathcal G^*(\vq,\omega)=\mathcal G(-\vq,-\omega)$ in Eq.~\eqref{eq:force_density_perturbation} isolates $\Im\mathcal G=(\mathcal G-\mathcal G^*)/(2\ii)$}. 
We refer to this factorization as the trajectory-response relation.

\begin{figure}[t]
\centering
\includegraphics[width=\columnwidth]{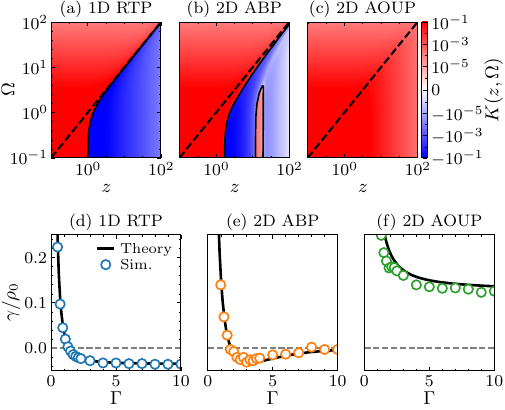}
\caption{
\textbf{Finite-frequency response and linear friction of active baths.}
(a-c) Dimensionless responses $K(z,\Omega)$ for (a) a one-dimensional RTP, (b) a two-dimensional ABP, and (c) a two-dimensional AOUP, with $z=qv_0\tau$ and $\Omega=\omega\tau$.
Color denotes the response on a symmetric logarithmic scale, and solid contours mark $K=0$.
The dashed line $\Omega=z$ corresponds to $V=v_0$ with $\vq\parallel\vV$.
(d-f) Corresponding linear friction coefficients $\gamma/\rho_0$ for a repulsive Gaussian probe as a function of $\Gamma$.
Solid curves are the weak-potential prediction in the $V\to0$ limit, and
open symbols show fixed-velocity simulations at $V/v_0=0.01$.
Horizontal dashed lines indicate zero friction.}
\label{fig:drag_vs_gamma}
\label{fig:matched_finite_frequency_kernels}
\end{figure}

\begin{figure*}[t]
\centering
\includegraphics[width=\textwidth]{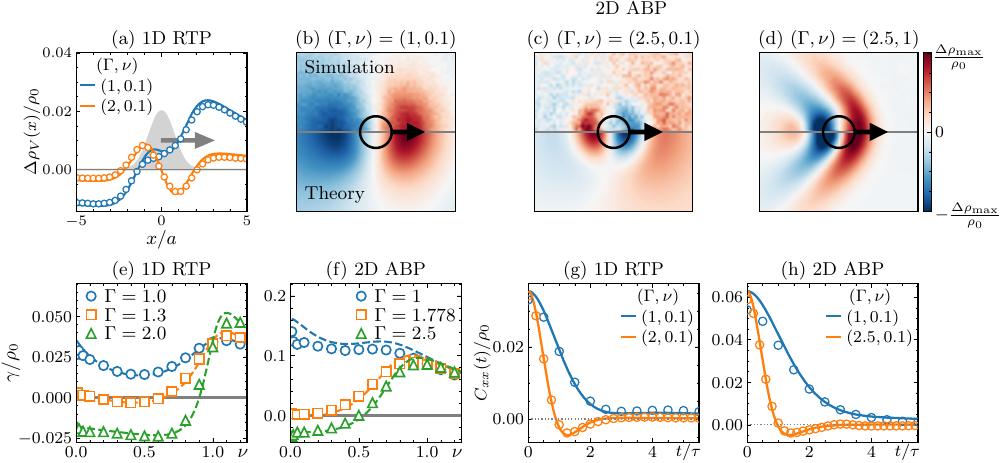}
\caption{
\textbf{Density profile, friction, and force autocorrelations.}
(a-d) Motion-induced density difference $\Delta\rho(\mathbf r, \nu)/\rho_0 = [\rho(\mathbf r,\nu)-\rho(\mathbf r,0)]/\rho_0$ for (a) 1D RTPs at $(\Gamma,\nu)=(1,0.1),(2,0.1)$ and (b-d) 2D ABPs at $(1,0.1)$, $(2.5,0.1)$, and $(2.5,1)$, where $\nu=V/v_0$.
In (a), curves represent theoretical prediction and symbols denote simulations.
In (b-d), the upper and lower half-panels show simulation and theory, respectively.
The shaded region and circles indicate the Gaussian probe with $a=1$, and arrows denote its direction of motion.
Each plot has an independent color scale with its own $\Delta\rho_{\max}/\rho_0 = \max\Delta\rho(\mathbf r,\nu)/\rho_0$.
The values of $\Delta\rho_{\max}/\rho_0$ are $0.025$, $0.0065$, $0.13$ for (b), (c), and (d), respectively.
(e,f) Nonlinear friction $\gamma(V)/\rho_0 = -F_x/(\rho_0V)$ for (e) 1D RTPs and (f) 2D ABPs.
Horizontal lines represent zero friction.
(g,h) Connected force autocorrelation $C_{xx}(t)/\rho_0 = \langle\delta F_x(0)\delta F_x(t)\rangle/\rho_0$ for (g) RTPs and (h) ABPs.
In (e-h), curves represent theoretical predictions and symbols denote simulations. 
}
\label{fig:FiniteV}
\end{figure*}

\textit{Classification of mechanical power transfer---}
For an inversion-symmetric free process, $\Phi_0(-\vq,t)=\Phi_0(\vq,t)=\Phi_0(\vq,t)^*$, so $\Im\mathcal G$ is odd in frequency.
Define
\begin{equation}
\mathcal K(\vq,\omega)=\frac{\Im\mathcal G(\vq,\omega)}{\omega},
\label{eq:mode_kernel}
\end{equation}
with its value at $\omega=0$ defined by continuity.
The mean power transferred from bath to probe is
\begin{equation}
\begin{aligned}
\dot W_{\rm bath}(\vV)&=\vV\cdot\vF_{\rm bath}(\vV)\\
&=-\mu\rho_0\int_{\vq}q^2|U_{\vq}|^2\omega_{\vq}^2
\mathcal K(\vq,\omega_{\vq}).
\end{aligned}
\label{eq:power_kernel}
\end{equation}
All factors other than $\mathcal K$ are nonnegative; hence
\begin{equation}
\mathcal K(\vq,\omega)\geq0\ \forall(\vq,\omega)
\quad\Longrightarrow\quad\dot W_{\rm bath}(\vV)\leq0
\label{eq:modewise_criterion}
\end{equation}
for every imposed velocity and every weak probe.
Negative-$\mathcal K$ modes permit positive-power contributions, but reversal of the total drag for a particular probe depends on its weight $|U_{\vq}|^2$.

We compare active Ornstein--Uhlenbeck particles (AOUPs), RTPs, and ABPs with matched velocity correlations~\cite{shaebani2020computational}:
\begin{equation}
\avg{u_i(t)u_j(0)}=\frac{v_0^2}{d}\ee^{-\abs{t}/\tau}\delta_{ij}.
\label{eq:matched}
\end{equation}
They share the root-mean-square speed $v_0$, persistence time $\tau$, and long-time diffusivity $D_a=v_0^2\tau/d$~\cite{cates2013when}.
Previous studies have shown that similar low-order statistics need not imply identical responses or energetics~\cite{solon2015active,hancock2015effect,lee2022effects}.
Equation~\eqref{eq:trajectory_response_characteristic} illustrates a related distinction: the shared velocity correlation fixes the second cumulant of the displacement, whereas the self-ISF retains higher-order displacement statistics that can distinguish the response spectra~\cite{martens2012probability,kurzthaler2016intermediate,malakar2018steady,basu2018active,basu2019long-time}.
We first set $D_t=0$.
For these isotropic models, the response depends on $\vq$ only through $q=|\vq|$.
With the dimensionless variables $z=qv_0\tau$ and $\Omega=\omega\tau$, we write $\mathcal K(q,\omega)=\tau^2K(z,\Omega)$.
Detailed derivations are given in the SM~\cite{prlsm}.

For AOUPs~\cite{fodor2016how, martin2021statistical, bonilla2019active}, Gaussian displacements give
\begin{equation}
\Phi_0(q,t)=\exp\!\left[-\frac{z^2}{d}\left(\frac{t}{\tau}-1+\ee^{-t/\tau}\right)\right].
\label{eq:aoup_self_isf}
\end{equation}
This self-ISF is positive and monotonically decreasing for each $q>0$, and its sine transform gives $K(z,\Omega)\geq0$ in every dimension.
Equation~\eqref{eq:modewise_criterion} excludes positive mechanical power at every speed, for any weak probe.

A complete-reset RTP~\cite{tailleur2008statistical,martens2012probability,mori2020universal,angelani2015run-and-tumble,malakar2018steady} moves with $\vu=v_0\vn$, redrawing $\vn$ uniformly on $S^{d-1}$ at rate $\alpha=\tau^{-1}$.
In $d=1$, negative response occurs when
\begin{equation}
z^2>1+\Omega^2.
\label{eq:rtp1_negative_condition}
\end{equation}
Uniform translation samples $\Omega=(V/v_0)z$, so these modes are accessible only for $|V|<v_0$.
In contrast, for $d\geq2$, $K(z,\Omega)\geq0$ at all wave numbers and frequencies, excluding positive steady power transfer to any weak probe.

For ABPs~\cite{romanczuk2012active,sevilla2015smoluchowski}, $\vu=v_0\vn$ and $\vn$ undergoes rotational diffusion with $\tau=[(d-1)D_r]^{-1}$.
We evaluate the self-ISF through its angular spectral representation~\cite{kurzthaler2018probing,kurzthaler2016intermediate}.
In the static limit, the leading short-wavelength response is $(d-2)/z^2$ for $d\geq3$.
In $d=2$ this ballistic contribution cancels, and the first rotational correction gives
\begin{equation}
K(z,0)=-\frac{1}{2z^3}+\order(z^{-4}).
\label{eq:abp_2d_large_z_boxed}
\end{equation}
Thus two-dimensional ABPs have negative static response at sufficiently short wavelengths.
At long wavelengths, diffusive spreading gives $K(z,0)>0$.
For $d\geq3$, numerical and large-$d$ asymptotic results support the conjecture of nonnegative ABP response at all wave numbers and frequencies.

The responses $K$ for representative models and dimensions are summarized in Fig.~\ref{fig:matched_finite_frequency_kernels} (a)-(c).
Motion samples $\Omega=(V/v_0)z\cos\varphi$, where $\varphi$ is the angle between $\vq$ and $\vV$.
In one dimension this fixes $|\Omega|=(|V|/v_0)z$; in $d\geq2$, oblique wave vectors can sample smaller frequencies.
Access to a negative-response sector does not by itself imply positive net power: its contribution still carries both the interaction weight and the factor $\omega_{\vq}^2$ in Eq.~\eqref{eq:power_kernel}.

The all-speed exclusions above survive arbitrary independent translational diffusion, which multiplies the self-ISF by $\ee^{-D_tq^2t}$ and induces a positivity-preserving transform of $\mathcal K$.

\textit{Linear friction for a specified probe---}
For an isotropic bath and probe, write $\vF_{\rm bath}=F_\parallel(V)\widehat{\vV}$ and $\gamma(V)=-F_\parallel(V)/V$ for $V=|\vV|>0$.
The small-speed response $F_{{\rm bath},i}=-\gamma_{ij}V_j+\order(V^2)$ follows from Eq.~\eqref{eq:finiteV_general}:
\begin{equation}
\gamma_{ij}=\mu\rho_0\int_{\vq}q_iq_jq^2|U_{\vq}|^2\mathcal K(\vq,0).
\label{eq:generaldrag_tensor}
\end{equation}
We choose the repulsive Gaussian potential $U_g(r)=\epsilon\exp[-r^2/(2a^2)]$, with $\epsilon>0$.
The ratio $\Gamma=v_0\tau/a$ compares the persistence length with the probe width.

For 1D RTPs, the Gaussian-probe friction changes sign at $\Gamma_c=\sqrt2$.
Below this value, the positive long-wavelength contribution dominates; above it, the interaction weights negative-response modes sufficiently strongly to reverse the total friction.
For 2D ABPs the corresponding crossing is at $\Gamma_c\simeq1.78$.
By contrast, the AOUP response is positive at every wave number, so changing $\Gamma$ cannot reverse its linear friction.
Direct fixed-velocity particle simulations reproduce these predictions [Fig.~\ref{fig:drag_vs_gamma} (d)-(f)].

\textit{Motion-induced density distortion and force---}
The density field provides a spatial interpretation of the force reversal.
However, the sign of the net force cannot be inferred from a local density accumulation behind the probe alone; it depends on the full density distortion weighted by $\nabla U$~\cite{demery2014generalized,peng2022forced,knezevic2021oscillatory}.
To isolate the change caused by translation, we define $\Delta\rho(\br,V)=\rho(\br,V)-\rho(\br,0)$ by subtracting the resting-probe profile.
For an inversion-symmetric bath and probe, this reference profile is inversion-even and contributes no net force.
The trajectory-response relation then gives
\begin{equation}
\Delta\rho_{\vq}(V)=-\mu\rho_0q^2U_{\vq}
\left[\mathcal G(\vq,\omega_{\vq})-\mathcal G(\vq,0)\right].
\label{eq:motion_induced_density_fourier}
\end{equation}
For the repulsive Gaussian potential, density enhancement ahead of the probe contributes resistance, whereas enhancement behind it contributes propulsion.

Figure~\ref{fig:FiniteV}(a--d) illustrates how this distortion changes with persistence and speed.
At $V/v_0=0.1$ and $\Gamma=1$, both the one-dimensional RTP and the two-dimensional ABP show a drag-producing density contrast, with enhancement ahead of the probe and depletion behind it.
At larger persistence ($\Gamma=2$ for RTPs and $\Gamma=2.5$ for ABPs), the contrast reverses near the probe and the force becomes propulsive, even though the density farther ahead can remain enhanced.
Increasing the speed can restore resistance: for the ABP at $\Gamma=2.5$ and $V=v_0$, density enhancement reappears ahead of the probe.
The inverse Fourier transform of Eq.~\eqref{eq:motion_induced_density_fourier} quantitatively reproduces both the spatial form and the amplitude of these simulated profiles.

The friction curves in Fig.~\ref{fig:FiniteV}(e,f) quantify this reversal and the recovery of positive friction, in agreement with simulations for both RTPs and ABPs.
As the probe speed changes, the interaction weight remains fixed while the imposed motion selects different $\omega_{\vq}$ of the same bath response.
This frequency selection accounts for the velocity dependence of the mean force.

Such a velocity-dependent mean force can support spontaneous motion of a free passive probe and select its steady speed~\cite{kim2024symmetry-breaking,pei2026transfer}.
The resulting free-probe dynamics, together with the general recovery of resistance at sufficiently large speeds for localized weak potentials, are analyzed in the companion paper~\cite{companionPRE}.

\textit{Force fluctuations---}
Beyond the mean force, the same free trajectories determine force fluctuations at order $U^2$.
For nonnegative time lag $t$, the stationary connected covariance is
\begin{equation}
\begin{aligned}
C_{ij}(t;\vV)
&=\rho_0\int\dd^dr\, \bigl\langle\partial_iU(\br+\Delta\br_t)\partial_jU(\br)\bigr\rangle\\
&=\rho_0\int_{\vq}q_iq_j|U_{\vq}|^2 \Re\!\left[\ee^{\ii\omega_{\vq}t}\Phi_0(\vq,t)\right].
\end{aligned}
\label{eq:force_covariance_trajectory}
\end{equation}
At equal times, $C_{ij}(0;\vV)$ is independent of both $\vV$ and the bath's propulsion dynamics, so the leading propulsion dependence enters through the subsequent time evolution.
The predicted force autocorrelations also agree with simulations~[Fig.~\ref{fig:FiniteV}(g,h)].

For the ideal active bath, the mean force and force covariance satisfy the direct relation
\begin{equation}
F_{{\rm bath},i}(\vV)=\mu\sum_j\int_0^\infty\frac{\dd t}{t}\frac{\partial C_{ij}(t;\vV)}{\partial V_j}.
\label{eq:force_covariance_relation}
\end{equation}
The equilibrium Green--Kubo relation connects linear friction to force autocorrelations~\cite{kubo1966fluctuation}, while a related time-integrated correlation determines the drag on inclusions moving through dissipative Gaussian fields~\cite{demery2010drag1,demery2010drag2}.
Unlike the equilibrium Brownian case, where free-particle displacement statistics are fixed by the diffusion coefficient, the mean force and covariance linked by Eq.~\eqref{eq:force_covariance_relation} depend on the full displacement statistics of active-particle trajectories~\footnote{In the equilibrium Brownian limit, Eq.~\eqref{eq:force_covariance_relation} recovers the Green--Kubo relation for linear friction~\cite{prlsm}.}.

\textit{Discussion---}
Space--time correlations have long connected microscopic dynamics to macroscopic observables~\cite{van1954correlations}.
Force calculations for uniformly moving perturbations separate medium dynamics from the spatial form of the perturbation~\cite{demery2010drag1,demery2010drag2,astrakharchik2004heavy,cherny2012superfluidity}.
Here, we extend the equilibrium connection between bath-particle displacement statistics and mechanical response to ideal active baths in the weak-probe limit.

The present results apply to an ideal bath of noninteracting active particles and weak probe--particle interactions, with the mean force calculated to order $\order(\rho_0 U^2)$.
This treatment excludes the impenetrable-probe limit~\cite{burkholder2020nonlinear,peng2022forced} and collective effects arising from interactions among bath particles~\cite{marchetti2013hydrodynamics,granek2020bodies,granek2024colloquium}.
For time-dependent probe motion, including oscillatory driving, the delayed response of the bath may also need to be accounted for~\cite{knezevic2021oscillatory}.

Within the dilute regime, the trajectory-response relation applies equally to other propulsion processes, with the corresponding response determined by their free-particle displacement statistics~\cite{kurzthaler2016intermediate,sevilla2021generalized,zhao2024quantitative,kurzthaler2024characterization,santra2020run-and-tumble,sevilla2020two,rusch2024intermediate}.
The same weak-coupling approach could be extended to combined translation and rotation, including probes in dilute chiral baths~\cite{liebchen2022chiral,poggioli2023odd,hargus2025odd}: 
for an orientable probe with interaction $U(\br,\theta)$, the conjugate derivative $-\partial_\theta U$ gives the bath torque.

Experimentally, a translating optical potential can realize a weak penetrable probe~\cite{evers2013particle,buttinoni2022active}.
Probe-free displacement statistics measured by single-particle tracking or differential dynamic microscopy~\cite{cerbino2008differential,wilson2011differential,kurzthaler2018probing,zhao2024quantitative,kurzthaler2024characterization} can then be used to predict the force for different probe sizes, shapes, and imposed velocities.
Conversely, probe measurements at different sizes, shapes, and velocities can constrain the wave-vector and frequency dependence of the bath response.

\section*{acknowledgments}
We thank Jae Dong Noh for helpful discussions.  
%This work was supported by [---]. 
%The numerical simulations were performed using computational resources provided by [---].  

%\section*{Data Availability}
%Code for the numerical evaluation of the theory presented in this work is available at [---].
%The code used for the numerical evaluation of the theory presented in this work will be made publicly available upon publication.

\bibliography{main}
\end{document}